\documentclass[prl,twocolumn,superscriptaddress]{revtex4-2}
\usepackage{bbm}
\usepackage{amsfonts}
\usepackage{amssymb}
\usepackage{amsmath}
\usepackage{graphicx}
\usepackage{dcolumn}
\usepackage{hhline} 
\usepackage{appendix}
\usepackage{bm}
\usepackage{float}
\usepackage{units}
\usepackage{txfonts}
\usepackage{makecell}
\usepackage{multirow}
\usepackage{CJKutf8}
\usepackage{subfigure}
\usepackage{color}
\usepackage{soul}
\usepackage{ulem}
\usepackage{url}
\usepackage[colorlinks,linkcolor=blue,anchorcolor=blue,citecolor=blue,urlcolor=blue]{hyperref}
\usepackage{array}
\begin{document}
\title{Reversible Steady Domain-Wall Motion Driven by a Direct Current}


\author{K. Y. Jing}
\affiliation{Institute for Advanced Study in Physics, Zhejiang University, 310027 Hangzhou, China}
\author{X. R. Wang}
\email[Contact author: ]{phxwan@cuhk.edu.cn}
\affiliation{School of Science and Engineering, Chinese University of Hong Kong (Shenzhen), Shenzhen, 51817, China}
\author{H. Y. Yuan}
\email[Contact author: ]{hyyuan@zju.edu.cn}
\affiliation{Institute for Advanced Study in Physics, Zhejiang University, 310027 Hangzhou, China}

\date{\today}

\begin{abstract}
Understanding and manipulating nanoscale domain wall (DW) dynamics is a central 
topic in magnetism and spintronics for its promising applications in logic and 
memory devices. In most magnetic systems, inertia affects only transient DW dynamics, while the long-time DW motion is uniquely determined by the magnitude and direction of the applied current. Here we show that this paradigm breaks down in ferrimagnets near the angular momentum compensation point. We demonstrate that a DW can propagate steadily either forward or backward even under a direct current, with the direction controlled solely by the current strength. 
This anomalous phenomenon originates from the inertial dynamics of an internal DW collective coordinate, which behaves as a massive object evolving in a current-dependent double-well potential. Depending on the driving current, the system relaxes into distinct stable states associated with opposite directions of motion. Our findings reveal an unexpected role of inertia in nonlinear spin dynamics, and enable low-energy spintronic functionalities including sensitive magnetic-field detection and reconfigurable one-port devices.
\end{abstract}

\maketitle

{\it Introduction}---Magnetic domain walls (DWs) are fundamental excitations in ordered magnets and constitute versatile information carriers for logic and memory devices. Domain wall dynamics can be efficiently manipulated by a variety of stimuli, including magnetic fields \cite{FIMexper1,FIMexper6,FerriAC, FIMexper7,KYField1,ivanov2020}, 
electric currents \cite{FIMexper4,FIMexper9,FIMexper2,FIMexper3,yang2015,
FIMexper5,FerriCD1,Eduardo1,Ferrisc,FIMexper8,FerriSOT,KYcurrent1}, 
spin waves \cite{FerriSW,liang2022,yu2018,lan2017,yan2011}, thermal gradient \cite{Ferrithermal,xiansi2}, 
and material inhomogeneity \cite{Ferrianisotropy}.
Among these, current-induced spin-orbit torques (SOTs) are particularly attractive due to 
their efficiency and compatibility with CMOS structures \cite{Parkin1}.
Conventionally, ferromagnetic DW dynamics was studied, whose motion has 
a distinguished breakdown field, beyond which the DW velocity ceases to increase. 
To circumvent this issue, antiferromagnets and ferrimagnets were proposed to establish a 
high-speed and stable DW motion. In particular, ferrimagnets have the merit that they
combine the advantages of ferromagnets with easy manipulation and 
antiferromagnets with ultrafast spin dynamics \cite{Review1}.
Also, ferrimagnets have much wider tunability compared to ferromagnetic and 
antiferromagnetic materials \cite{Review1,zhang2023}.




Magnetization dynamics in a ferromagnet is described by 
a first-order nonlinear partial differential equation (PDE) called 
Landau-Lifshitz-Gilbert (LLG) equation \cite{landau1935,gilbert}, 
thus spin has no  inertial effect \cite{xrw1,xrw2} with only a few exceptions in high-frequency 
regime \cite{quarenta2024,yuan2025,neeraj2021}. 
The situation changes in antiferromagnets and ferrimagnets, which are described by 
two coupled LLG equations. Then the resulting equation of the N{\'e}el order 
defined as the magnetization difference of two sublattices has a second order 
derivative of time, generating a non-vanishing inertia of magnetic structures, 
such as DWs, vortices, and skyrmions \cite{FerriAC,kim2017vortex,kim2017skyrmion,liu2022}. 
It is widely believed that inertia
affects only transient DW behavior on the time scale of 
picosecond to sub-nanosecond, while the steady DW velocity under a current is uniquely determined by the current direction
and magnitude \cite{effLLG4,Eduardo1,Ferrisc,shiino2016,tveten2013}. 
Reversing the propagation direction therefore requires reversing the current polarity. 
This understanding has shaped
the conventional view of current-driven DW dynamics.

\begin{figure}
\centering
\includegraphics[width=8.5cm]{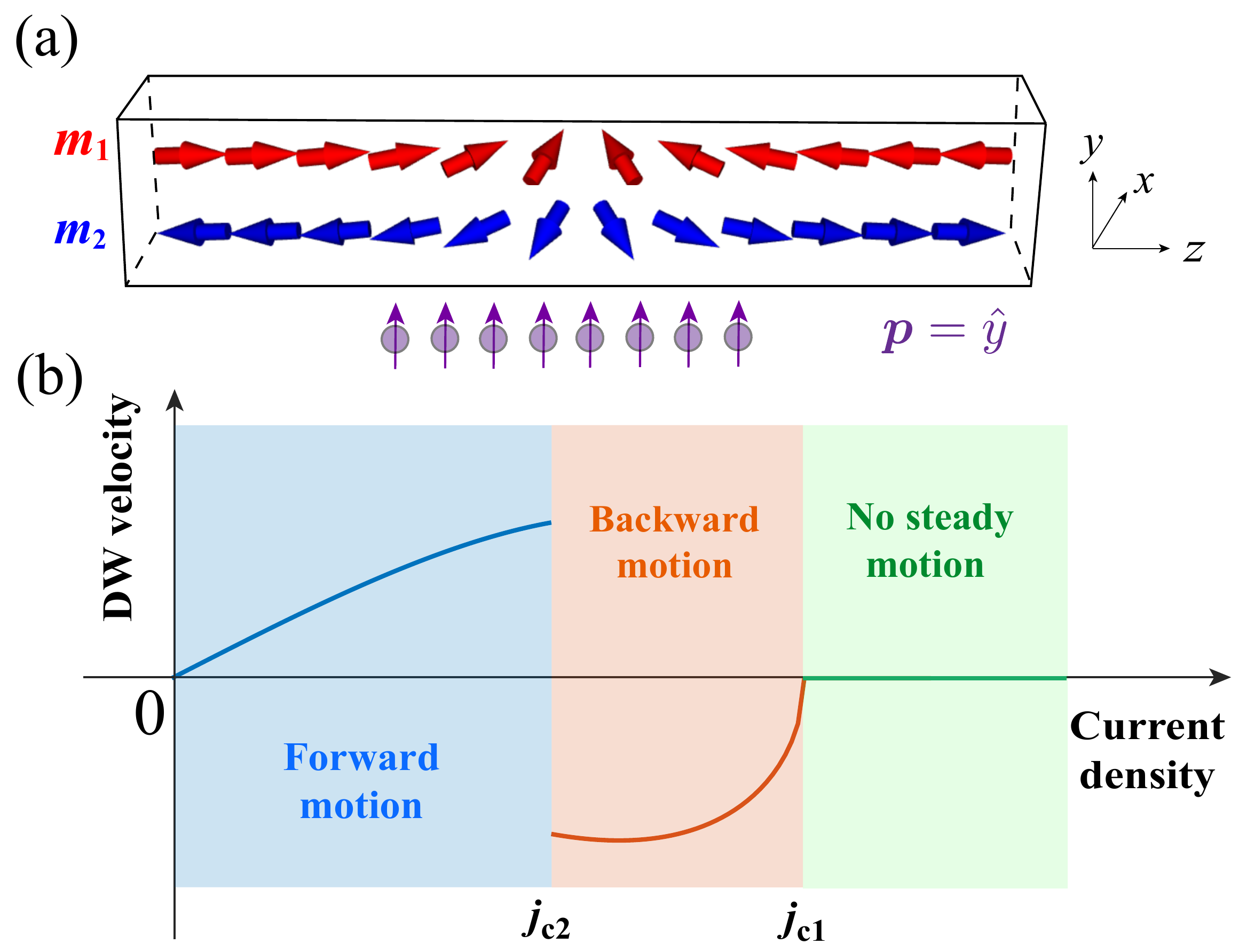}\\
\caption{
(a) Schematic of a head-to-head ferrimagnetic DW. 
(b) Schematic of the steady DW velocity driven by a direct current. 
The blue, orange, and green
regions indicate the three different regimes of DW motion.
$j_{\mathrm{c1}},j_{\mathrm{c2}}$ are two critical current densities.
}
\label{schem}
\end{figure}

In this letter, we show that this widely accepted picture can break down under certain conditions.
Specifically, in a ferrimagnet near the angular momentum compensation point (AMCP), inertia qualitatively 
alters steady DW dynamics and enables a striking phenomenon: the direction of steady DW motion can be reversed under
a direct current (DC) without changing the current polarity, as 
summarized schematically in Fig. \ref{schem}. 
This behavior arises from the inertial dynamics of the 
DW tilting angle, which experiences a double-well potential shaped
by magnetic anisotropy and spin torque. Depending on the current amplitude, 
the DW relaxes into different stable states,
corresponding to opposite propagation directions. Consequently, the steady 
DW velocity exhibits a reversal at
a critical current, leading to a bistable dynamical response. The physical insight 
can be extended beyond magnetism and is generally applicable to a broad class 
of driven-dissipative systems with inertial degrees of freedom.  
From an application perspective, the bistable DW motion driven by direct current enables the design of novel 
spintronic devices including sensitive magnetic field detection and reconfigurable one-port devices.

{\it Model and method.}---
We consider a head-to-head (HH) ferrimagnetic DW consisting of two 
antiferromagnetically coupled sublattices, as shown in Fig. \ref{schem}(a), 
where $\boldsymbol{m}_1,\boldsymbol{m}_2$
are unit vectors of the magnetizations in each sublattice.
Spin current with spin polarization direction $\boldsymbol{p}=\hat{y}$ 
(purple arrows in Fig. \ref{schem}(a)) 
is injected to the ferrimagnet, exerting spin torques on the local magnetizations.
For the sake of generality, we do not specify the physical origin of the spin torque.
It can be originated from
spin accumulation arising at the ferrimagnet-heavy metal interface due to spin-orbit coupling 
effects (spin-orbit torques), such as spin Hall effect, 
Rashba effect or other mechanisms \cite{Song2021}, or 
from the direct spin injection (spin transfer torque) in a spin valve or magnetic 
tunneling junction structure \cite{saidaoui2014,abert2017,berger1996,Slonczewski}.
Our model can even be extended to recently reported orbital torques
which take the same mathematical form \cite{wang2024,yang2024,fukami2025}.

The ferrimagnetic magnetization dynamics in the presence of spin torques is 
governed by two coupled LLG equations \cite{Yuan1,AFM1}
\begin{subequations}\label{dynamics}
\begin{align}
 {\partial_t \boldsymbol{m}_{{1}}}& \!  = \! -\boldsymbol{m}_{{1}}  \! \times
\left(- \frac{1}{s_1}\frac{\delta E}{\delta \boldsymbol{m}_1} \! -\! \alpha_{11} 
 {\partial_t \boldsymbol{m}_{{1}}}
\!-\! \alpha_{12}  {\partial_t \boldsymbol{m}_{{2}}}
 \right) \!+\!  \gamma_1 \boldsymbol{\tau}_{{1}},    \!
 \\ {\partial_t \boldsymbol{m}_{{2}}}& \! = \! 
-\boldsymbol{m}_{{2}} \! \times
\left(-\frac{1}{s_2}\frac{\delta E}{ \delta \boldsymbol{m}_2}  \! - \! \alpha_{22}  
{\partial_t\boldsymbol{m}_{{2}}}
\!-\!\alpha_{21}  {\partial_t \boldsymbol{m}_{{1}}}
\right) \!+\! \gamma_2 \boldsymbol{\tau}_{{2}}, \!
\end{align}
\end{subequations}
where $s_\ell = M_\ell /\gamma_\ell$, $M_\ell$, $\gamma_\ell$ are the spin densities, saturation 
magnetizations, and gyromagnetic ratios in $\ell$-th sublattice, 
$\alpha_{11},\alpha_{22}$ and $\alpha_{12},\alpha_{21}$ are the intrasublattice
and intersublattice damping coefficients, respectively.
$E=\int J \boldsymbol{m}_1 \cdot\boldsymbol{m}_2+
\sum_{\ell=1,2}\left[A_\ell(\nabla\boldsymbol{m}_\ell)^2+f_\ell(\boldsymbol
{m}_\ell)\right] d^3 \boldsymbol{x}$ is the 
total magnetic energy with $J$, $A_\ell$, and $f_\ell$ 
being the intersublattice coupling, intrasublattice exchange stiffness, and
magnetic anisotropy, respectively.
The spin torque acting on the magnetization includes damping-like and field-like 
components, 
i.e. $\tau_\ell
=- \mu_0 [H_{\mathrm{DL},\ell} 
\boldsymbol{m}_\ell \times ( \boldsymbol{m}_\ell \times \boldsymbol{p} ) +
H_{\mathrm{FL},\ell} \boldsymbol{m}_\ell \times \boldsymbol{p} ] $ (with $\mu_0$ being the
vacuum permeability), 
and their relative strength
is characterized by the ratio $\xi_\ell = H_{\mathrm{FL},\ell}/ H_{\mathrm{DL},\ell}$.
In general, the spin torque coefficients
$H_{\mathrm{DL},\ell}$ and $H_{\mathrm{FL},\ell}$ are both proportional to 
the applied current density. 
Throughout this letter, we assume a positively flowing current
such that $H_{\mathrm{DL},\ell} >0 $ and $H_{\mathrm{FL},\ell} > 0$.

When the strength of intersublattice coupling ($J$) dominates
other interactions and external stimuli, the magnetic moments of the two sublattices 
are approximately antiparallel to each other. Then it is convenient to investigate the 
low-energy dynamics of ferrimagnets by introducing the N{\'e}el vector 
$\boldsymbol{n}=(\boldsymbol{m}_1 - \boldsymbol{m}_2)/2$ and magnetization vector
$\boldsymbol{m}=(\boldsymbol{m}_1+ \boldsymbol{m}_2)/2$. After rewriting Eqs. \eqref{dynamics} 
in terms of $\boldsymbol{m}$ and $\boldsymbol{n}$ and considering $|\boldsymbol{m}| \ll 1$, one can 
eliminate the slave variable $\boldsymbol{m}$ and derive a self-consistent equation of 
$\boldsymbol{n}$ as \cite{Review1,rodriguez2024}
\begin{equation}\label{Neel}
-\frac{s^2}{4J}  \boldsymbol{n} \times \partial_{tt}{\boldsymbol{n}} +\delta_s 
 \partial_t{\boldsymbol{n}} =- \boldsymbol{n} \times \boldsymbol{h}_{\boldsymbol{n}}
+\alpha s \boldsymbol{n}  \times \partial_t{\boldsymbol{n}}
+\boldsymbol{\tau}_{\boldsymbol{n} },
\end{equation}
where $\boldsymbol{h}_{\boldsymbol{n}}=-\delta E/\delta \boldsymbol{n}$ is the effective field
conjugate to $\boldsymbol{n}$,
$s=s_1 + s_2$ is the total spin density, $\delta_s = s_1 - s_2$ is the net spin density, damping coefficient
$\alpha = (\alpha_{11}s_1 + \alpha_{22}s_2 - \alpha_{12}s_1 - \alpha_{21}s_2 )/s$.
The spin torque is $\boldsymbol{\tau}_{\boldsymbol{n}}= -\tau_{\mathrm{DL}} 
\boldsymbol{n}\times (\boldsymbol{n} \times \boldsymbol{p})
- \tau_{\mathrm{FL}} \boldsymbol{n} \times \boldsymbol{p}$, with
$\tau_{\mathrm{DL}}=\mu_0 M_1 H_{\mathrm{DL},1}+\mu_0 M_2 H_{\mathrm{DL},2} $,
$\tau_{\mathrm{FL}}=\mu_0 M_1 H_{\mathrm{FL},1} - \mu_0 M_2 H_{\mathrm{FL},2} $.

To explore DW dynamics, we consider a widely used \cite{Walker} biaxial anisotropy 
$f(\theta,\phi)=-K_z \cos^2 \theta 
+K_y \sin^2 \theta \sin^2 \phi$, then the DW profile can be approximated by the standard 
Walker ansatz \cite{Walker} as
$\theta(z,t)\simeq 2\arctan \{ \exp[(z-z_0 (t)) /\Delta] \}$, 
$\phi(z,t)\simeq \phi(t)$, where $\theta$ and $\phi$ are the polar and 
azimuthal angles of the N{\'e}el vector, $z_0$ and
$\Delta(\phi)=\sqrt{A/(K_z+K_y\sin^2 \phi)}$ 
 are respectively the position and width of the DW.  
Substituting the Walker profile into Eq. \eqref{Neel} and integrating it with
$\int dz \sin \theta \times {\text{Eq. }} \eqref{Neel}$, we can
rewrite the dynamic equations in the collective coordinates $(z_0,\phi)$ as
\begin{subequations}\label{collecteq}
\begin{align}
&\frac{s^2}{4J}  \frac{\partial_{tt}{z}_0 }{\Delta}+ \delta_s \partial_t{\phi} 
+  \frac{\alpha s}{ \Delta}\partial_t{z}_0
= \frac{\pi}{2}\tau_{\rm DL} \cos \phi,
\label{Nphicollect}
\\
& \frac{s^2}{4J} \partial_{tt}{\phi} - \frac{\delta_s }{\Delta}\partial_t{z}_0+\alpha s \partial_t{\phi}
=-  K_y \sin 2\phi    +\frac{\pi}{2}\tau_{\rm FL} \cos \phi. 
\label{Nthetacollect}
\end{align}
\end{subequations}

{\it Inertia-free DW motion.}--- In the limit of vanishing inertia ($J\to \infty$), the terms proportional 
to $\partial_{tt} z_0$, $\partial_{tt} \phi$ vanish in
Eqs. \eqref{collecteq}, and we have
\begin{equation}\label{LLGphiint}
\partial_{t} \phi=\frac{\cos \phi }{(\alpha s)^2+\delta_s^2}\left[
\left(\frac{\pi}{2}\alpha s \tau_{\rm FL}+\frac{\pi}{2} \delta_s \tau_{\rm DL}\right)
 -2\alpha s K_y \sin \phi  \right],
\end{equation}
\begin{equation}\label{LLGz0}
\partial_{t} z_0
= \frac{ \pi}{2\alpha s }\tau_{\rm DL} 
\Delta \cos \phi - \frac{ \delta_s}{\alpha s} \partial_{t} \phi \Delta.
\end{equation}

Note that the steady DW motion does not involve the precession of DW plane, 
i.e. $\partial_{t} \phi=0$ in Eq. \eqref{LLGphiint}, which allows us to determine the steady DW tilting angle as
$\sin \phi_f = \eta \equiv  \pi \left(\alpha s \tau_{\rm FL}+
\delta_s \tau_{\rm DL}\right)/(4\alpha s K_y)$ or $\cos \phi_f=0$.
By substituting the steady $\phi_f$ into Eq. \eqref{LLGz0}, we derive the long-term averaged DW velocity as $\langle \partial_t z_0\rangle
= \pi  \tau_{\rm DL} \Delta \cos \phi_f /(2\alpha s)$. For very strong current above a threshold $j_{\mathrm c1}$ such that $\eta \geq 1$, the DW tilting angle always relaxes to $\pi/2$ and the steady DW velocity is zero.
When $j < j_{\mathrm c1}$, Eq \eqref{LLGphiint} admits two stable points,
$\phi_L=\sin^{-1}(\eta)$ and $\phi_R=\pi-\sin^{-1}(\eta)$ with opposite DW velocities.
In the inertia-free limit, the DW tilting angle deterministically relaxes to
either $\phi_L$ or $\phi_R$ depending on the initial condition.
A finite inertia term will allow the mutual transition of DW between 
these two stable points, as we shall see below.

\begin{figure}
\centering
\includegraphics[width=8.5cm]{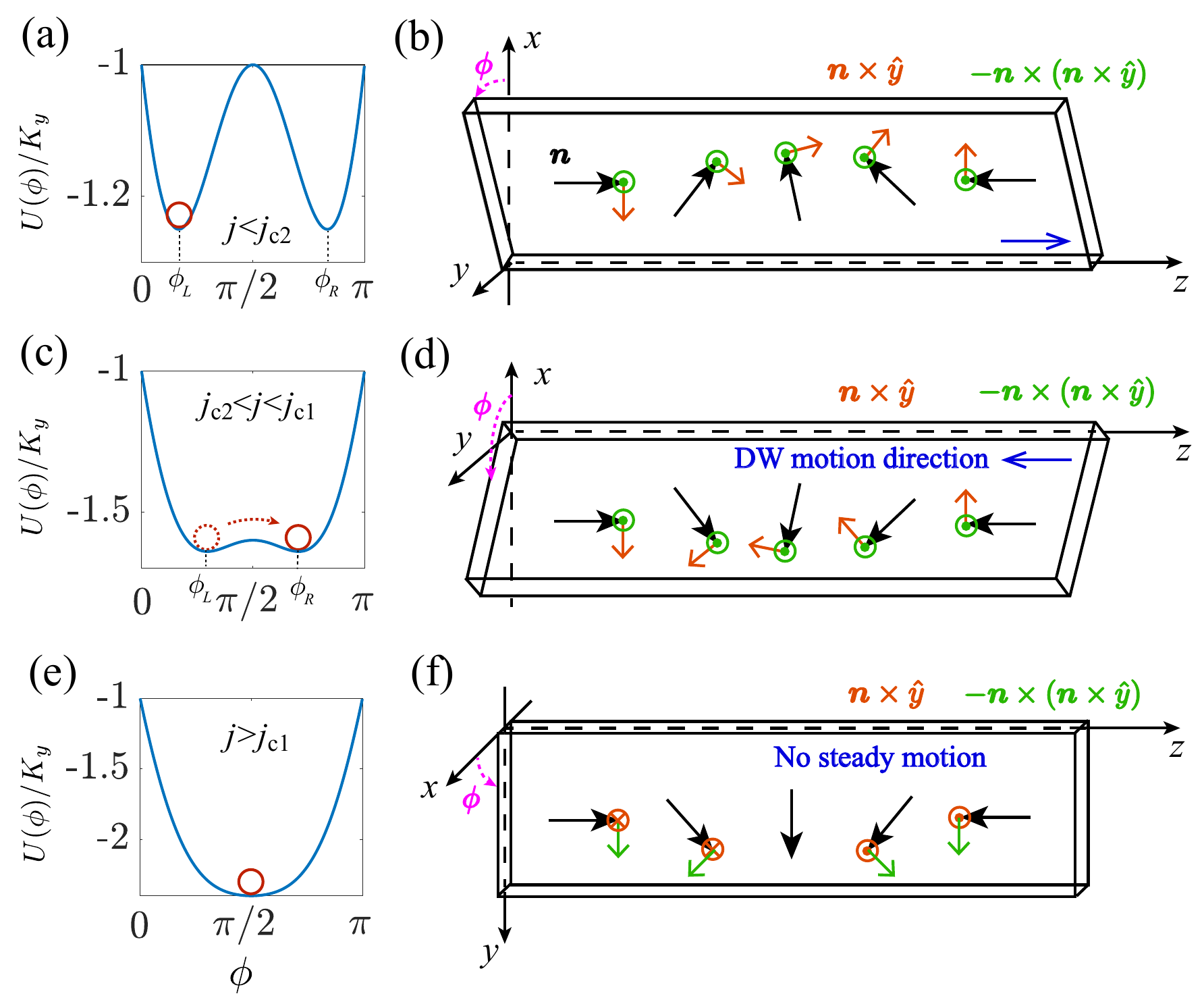}\\
\caption{Illustration of 
potential landscapes and torque analysis 
of a ferrimagnetic DW driven by spin torque for different current densities,
(a,b) $j<j_{\mathrm{c2}}$, 
(c,d) $j_{\mathrm{c2}}<j<j_{\mathrm{c1}}$,
(e,f) $j>j_{\mathrm{c1}}$.
The orange circles in (a,c,e) indicate the position of stable DW tilting angles,
the orange and green arrows indicate the directions of spin torques
of $\boldsymbol{n}\times\hat{y}$ and $-\boldsymbol{n}
\times(\boldsymbol{n}\times\hat{y})$.} 
\label{illus}
\end{figure}

{\it Inertia-driven DW motion.}--- To illustrate the inertia driven DW dynamics, we first consider a ferrimagnet 
at the AMCP for analytical simplicity. We emphasize that the physical mechanism discussed below remains valid away from the AMCP as we will prove later. By keeping the inertia terms in Eq. \eqref{Nthetacollect},  
the fully nonlinear dynamics of DW tilting angle is governed by 
\begin{equation}\label{phiinertial}
\frac{s^2}{4J} \partial_{tt} \phi =- \alpha s  \partial_t\phi -K_y \sin 2\phi 
+ \frac{\pi}{2}\tau_{\rm FL} \cos \phi.   
\end{equation}
This equation shows that the DW tilting angle $\phi$ behaves as a particle with an effective mass $s^2/(4J)$ moving in a potential
 $U(\phi)=-K_y \cos^2 \phi - \frac{\pi}{2}\tau_{\rm FL} \sin \phi$, and subject to a viscous damping force $-\alpha s  \partial_t \phi$.
The corresponding potential landscapes for different spin-torque strengths are plotted 
in Fig. \ref{illus}.
When $\eta=\pi \tau_{\rm FL} /(4 K_y)<1$ ($j<j_{\mathrm{c1}}$), 
the potential exhibits two local energy minima at $\phi_L$ 
and $\phi_R$, 
respectively, as shown in Figs. \ref{illus}(a,b). These minima are separated by an energy barrier of height 
$K_y (1-\eta)^2$, located at $\phi=\pi/2$. As the electric current, and hence $\eta$ increases, the energy 
barrier between the two minima decreases and eventually vanishes when 
$\eta \geq 1$, leaving a single minimum at $\phi=\pi/2$.

For a DW initially lying in the $xz$-plane ($\phi=0$), neglecting the inertial term implies that the DW 
always relaxes to the first local minimum it encounters, namely  $\phi_L\in(0,\pi/2)$. This leads to
a forward steady DW propagation with $\langle \partial_t z_0\rangle>0$.  
The situation changes qualitatively once inertia is taken into account. Now a second critical
current density $j_{\mathrm{c2}}(<j_{\mathrm{c1}})$ emerges, at which the energy barrier between the two local 
minima becomes sufficiently shallow. As a result, the DW can overcome the 
barrier via the inertial dynamics, and relaxes into the second 
minimum at $\pi - \sin^{-1}(\eta)$ (See Fig. \ref{illus}(c)).
Even though the two local minima have same value of $\sin \phi$, their 
$\cos \phi$ values have opposite signs, which directly implies a reversal of the DW moving direction 
according to Eq. \eqref{LLGz0}. 

The reversible DW motion can also be understood by analyzing the directions of spin torques acting
on the DW.
 For current density below $j_{\mathrm{c2}}$, the DW plane relaxes to a potential valley with $\phi \in (0,\pi/2)$,
as shown in Fig. \ref{illus}(a). In this regime, the spin torque along 
$\boldsymbol{n}\times (\boldsymbol{n} \times \hat{y})$  tends to increase the DW plane angle $\phi$, as indicated by green arrows in
Fig. \ref{illus}(b), while the torque arising from the hard-axis anisotropy counteracts this tendency by suppressing 
the increase of $n_y$. The balance between these two torques stabilizes the DW at $\phi=\phi_L$. 
Meanwhile, the spin torque along 
$\boldsymbol{n} \times \hat{y}$ 
collectively drives a coherent rotation of spins within the DW (orange arrows 
Fig. \ref{illus}(b), resulting in a forward DW motion
along the  $+\hat{z}$ direction.

For current density in the window $j_{\mathrm{c2}}<j<j_{\mathrm{c1}}$,
the inertial effects enables the DW to overcome the energy barrier separating the two potential minima, causing the DW plane angle to relax into the range $\phi \in (\pi/2,\pi)$, as shown in Fig. \ref{illus}(c). 
A similar torque analysis is still applicable, although 
the relative orientations between $\boldsymbol{n}$ and $\hat{y}$ have changed.
Once a steady motion of DW is established, the spin torque along 
$\boldsymbol{n}\times (\boldsymbol{n} \times
\hat{y})$ now tends to decrease the DW plane angle $\phi$,
(as shown by green arrows Fig. \ref{illus}(d)), competing with the hard-axis anisotropy torque that favors
alignment of $\phi$ toward $\phi$. Importantly, the sign of $\boldsymbol{n}\times \hat{y}$
is reversed in this configuration, resulting in a backward DW propagation along $-\hat{z}$ direction.
For current density larger than $j_{\mathrm c1}$, the potential possesses a single minimum at $\phi = \pi/2$,
where the spin torques inside the DW cancel each other (see Figs. \ref{illus}(e-f)). In this case, the DW structure is deformed slightly but no steady 
propagation occurs.

Alternatively, one can understand the anomalous DW dynamics from the energy perspective.
In the steady-motion regime, the total energy of the system remains constant in time, implying that the energy dissipation rate
through damping should be completely compensated by the work done by the spin torque.
A straightforward calculation shows that equalizing the energy dissipation rate, $-2\alpha s (\partial_t z_0)^2 /\Delta$, 
with the positive work done by the spin torque, $\pi \partial_t z_0 \tau_{\mathrm{DL}}\cos \phi$, yields the steady DW velocity $\langle \partial_t z_0 \rangle=  \pi  \tau_{\rm DL} \Delta \cos \phi_f /(2\alpha s)$, which is in agreement with the result obtained from the collective coordinate equations.

{\it Numerical verification.}---
To verify our analytical predictions, we numerically solve the coupled LLG equations \eqref{dynamics}
using MuMax3 package \cite{mumax3}.
A magnetic nanostrip with geometry $8\mathrm{\, nm} \times\, 
2\mathrm{\, nm}\times\,   2048 \mathrm{\, nm}
$ is divided into meshes of size
$1 \mathrm{\, nm}\times 1\mathrm{\, nm}\times 0.5 \mathrm{\, nm}$. 
To model a rare-earth-transition-metal alloy (such as GdFeCo, GdFe, GdCo) 
\cite{FIMexper1,FIMexper4,FIMexper6,parameter1,parameter2,parameter3,
parameter4,FIMexper2,Shen2024}, we use exchange constants 
$A_1 =A_2= 1.1\times 10^{-11}\mathrm{\, J} / \mathrm{m}$, 
and magnetic anisotropy $f_\ell=-K_{\ell,z} m^2_{\ell,z}+K_{\ell,y}m^2_{\ell,y}$ with $K_{1,z}=K_{2,z}=0.7\,\mathrm{MJ}/\mathrm{m}^3$,
$K_{1,y}=K_{2,y}=0.07\,\mathrm{MJ}/\mathrm{m}^3$. 
Damping coefficients are
$\alpha_{11}=\alpha_{22}=0.01$, $\alpha_{12}=\alpha_{21}=0$, 
the saturation magnetizations are $ M_1 =1010 \,\mathrm{kA}/\mathrm{m}$ and 
$ M_2 =900 \,\mathrm{kA}/\mathrm{m}$. Unless stated otherwise, the inter-sublattice exchange coupling
is chosen as $J=225 \, \mathrm{MJ}/\mathrm{m}^{3}$.
For the damping-like torque, we choose $H_{\mathrm{DL},1}=H_{\mathrm{DL},2}$.
To generate a sizable field-like torque, we set
$\xi_1 =10$, $\xi_2 =0$.
In terms of effective parameters defined in Eq. \eqref{Neel}, 
$A= 2.2\times 10^{-11}\, \mathrm{J} / 
\mathrm{m}$, $K_z=1.4 \,
\mathrm{MJ}/\mathrm{m}^3$,  and $K_y=0.14 \,
\mathrm{MJ}/\mathrm{m}^3$.
Initially, a DW lying in the $xz$ plane is prepared at the center of 
the nanowire, a spin torque is then applied to drive the DW motion and the DW velocity is obtained from 
the linear fit of time-evolution curve of the DW center where $m_z=0$. 

\begin{figure}
\centering
\includegraphics[width=8.5cm]{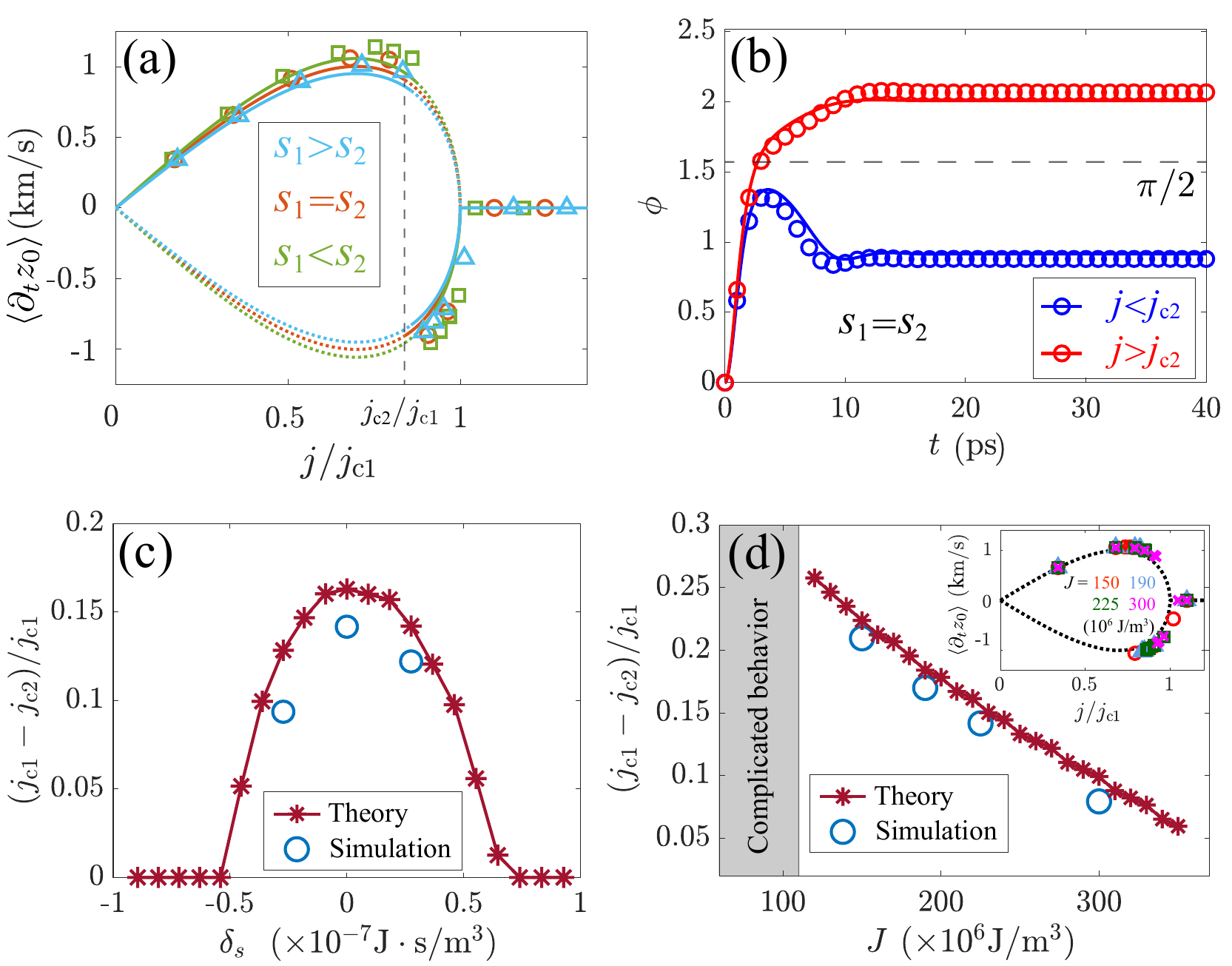}\\
\caption{(a) DW velocity as a function of applied current density in the vicinity of AMCP.
$\delta_s =0.275 \times10^{-7} \mathrm{J} \cdot \mathrm{s}/\mathrm{m}^3$ (light blue), 
$0$ (orange), $-0.272 \times10^{-7} \mathrm{J} 
\cdot \mathrm{s}/\mathrm{m}^3$ (light green).
Symbols are micromagnetic simulations and the curves are theoretical predictions
by Eq. \eqref{LLGz0}. The dashed line indicates the position of $j_{\mathrm{c2}}$. 
(b) The time evolution of DW plane angle for
$j/j_{\mathrm{c1}}=0.793$ (blue) and $0.907$ (red), respectively.
(c-d) The relative window width for reversible DW motion $(j_{\mathrm{c1}}-j_{\mathrm{c2}})
/j_{\mathrm{c1}}$ as a function of net spin density $\delta_s$ and inter-sublattice coupling $J$, respectively. 
The dark red stars directly connected by solid lines 
are numerical solutions to Eqs. \eqref{collecteq}, while the open blue circles
are micromagnetic simulations.} 
\label{vpyAMCP}
\end{figure}

Figure \ref{vpyAMCP}(a) shows the simulated DW velocity as a function of driving current
density for different $\delta_s$. The numerical results are in excellent agreement with the theoretical curves obtained by
solving Eqs. \eqref{collecteq}. 
Figure \ref{vpyAMCP}(b) presents the typical evolution of 
DW tilting angles $\phi(t)$ near $j_{\mathrm{c2}}$.
For currents slightly below $j_{\mathrm{c2}}$, 
$\phi$ finally relaxes to a value in $(0,\pi/2)$ (blue symbols).
For $j$ increases slightly above $j_{\mathrm{c2}}$,
the DW plane angle $\phi$ crosses the barrier at $\pi/2$ and ultimately relaxes to an angle in $(\pi/2,\pi)$.
This behavior is well captured by the numerical solution of Eq. \eqref{phiinertial} with proper initial conditions.

To examine the robustness of the reversible
DW motion driven by a direct current, we vary the tunable spin density $\delta_s$ and
plot the relative width for reversible DW motion, $(j_{\mathrm{c1}}-j_{\mathrm{c2}})/j_{\mathrm{c1}}$, as a function of $\delta_s$ in Fig. \ref{vpyAMCP}(c). Clearly, the current window width for reversible
DW motion is maximized near the AMCP and remains finite up to $\delta_s\sim 0.5 \times 10^{-7} \mathrm{J} \cdot \mathrm{s}/\mathrm{m}^3$. This value of spin density is feasible in experiments \cite{FIMexper1,FIMexper3,FIMexper4}.
Another critical parameter is the intersublattice antiferromagnetic
coupling strength $J$, which directly controls the DW inertia $s^2 /(4J)$.
Figure \ref{vpyAMCP}(d) shows the dependence of the reversal window on $J$, with the inset displaying the DW velocity as functions of current density for different coupling strength. As $J$ decreases, the DW has larger inertia to cross the energy barrier and thus enlarge the current window for reversible DW motion. 
As $J$ decreases below $110 \, \mathrm{MJ}
/\mathrm{m}^{3}$, the DW undergoes oscillations between
the two potential valleys at two sides of the $\phi=\pi/2$, 
and multiple reversal of DW propagation direction appears.
For even smaller values of $J$, the DWs in two sublattices
decouple, and the rigid-body motion of ferrimagnetic DWs does not exist.

\begin{figure}
\centering
\includegraphics[width=8.5cm]{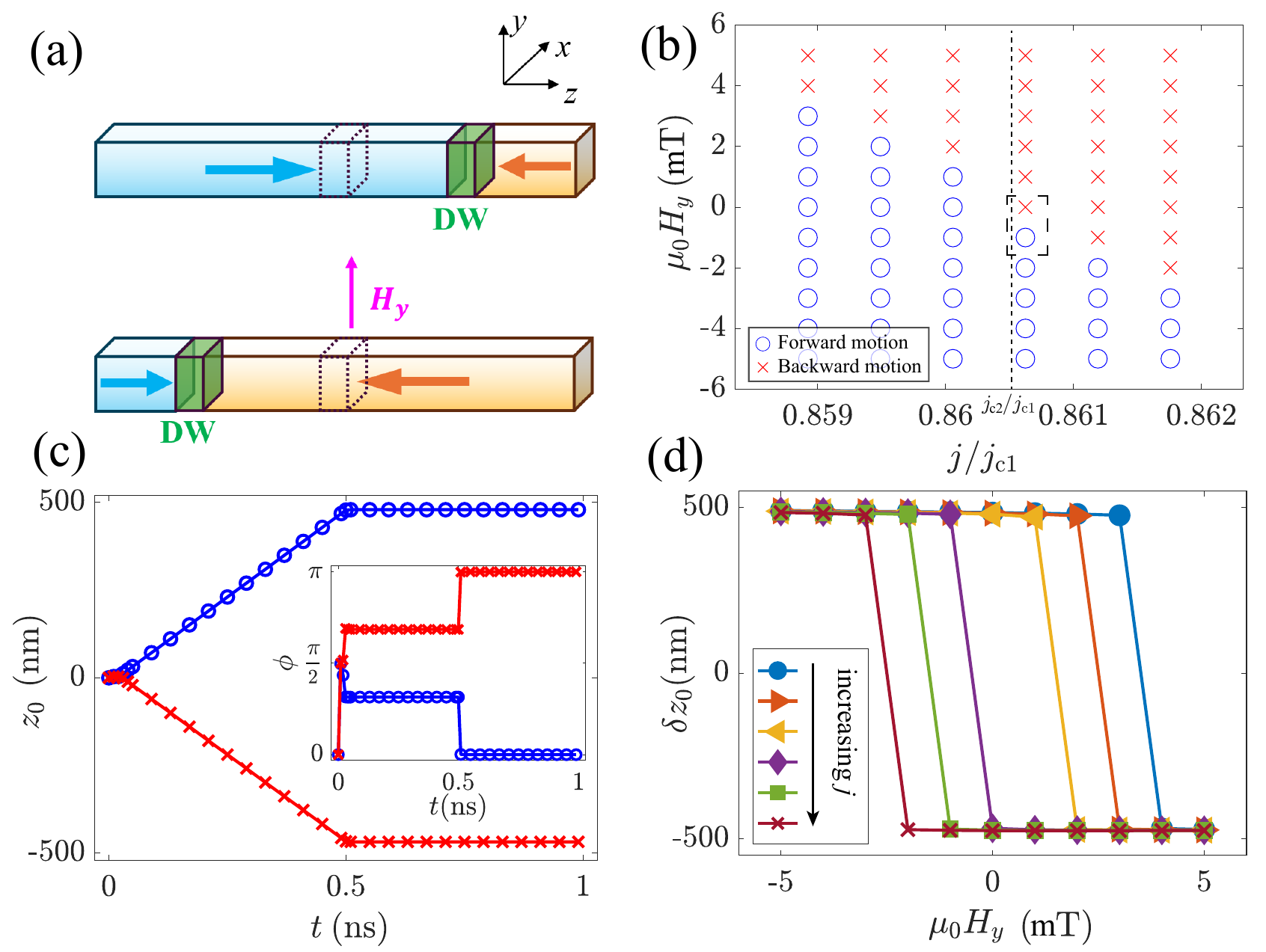}\\
\caption{(a)
Schematic of magnetic sensor based on DW motion.
(b) Direction of DW motion as a map of 
current density and 
external field.
(c) Two typical sets (boxed in (b)) of $z_0(t)$ and $\phi(t)$
for forward and backward motion.
(d) The DW displacement $\delta z_0$ after applying 
a direct current pulse with a duration of 0.5 ns,
the values of $j/j_{\mathrm{c1}}$ and $\mu_0 H_y$ are same as (b).
}
\label{sensor}
\end{figure}

{\it Magnetic sensor based on DW motion.}---The reversible DW motion driven by a direct current enables a sensitive magnetic sensor, which exploits the bistability of DW motion near the critical current $j_{\mathrm{c2}}$, as plotted
in Fig. \ref{sensor}(a). In this regime, the DW tilting angle evolves in a shallow double-well potential, allowing small perturbations to deterministically select one of two stable propagating directions. We apply a rectangular current pulse with a duration of 0.5 ns followed by an additional relaxation period of 0.5 ns. Fig. \ref{sensor}(b) shows
the direction of DW motion in the phase diagram of external magnetic 
field and applied current density. For a fixed current density,
the DW moving direction sensitively depends on the external field along the $y-$axis, which acts as an effective field-like spin torque on the DW tilting angle. A slight change in $H_y$ switches the DW between forward and backward motion, as shown in Fig. \ref{sensor}(c). By measuring the transition point between positive and negative DW displacement, as shown in Fig. \ref{sensor}(d), the external field along the $y-$axis can be determined with high sensitivity. Quantitatively, the DW direction reversal occurs when 
$j/j_{\mathrm{c1}}
+[\pi(M_1 - M_2)/(4K_y)] \mu_0 H_y=j_{\mathrm{c2}}/j_{\mathrm{c1}}
$. Accordingly, the minimum detectable field $\delta H_y$ is determined by the deviation of the applied current $\delta j$ from the critical value $j_{\mathrm{c2}}$
\begin{equation}
\mu_0 \delta H_y =-\frac{4\pi K_y}{ M_1 -M_2} \frac{\delta j}{j_{\mathrm{c1}}}.
\end{equation}
For our simulation $\delta j/j_{\mathrm{c1}}\simeq 6\times 10^{-4}$,
$ \mu_0 \delta H_y \simeq 1$ mT. Unlike conventional DW sensors that rely on continuous displacement, our approach converts a small magnetic field variation into a binary reversal of DW moving direction, providing a robust and fast sensing mechanism compatible with compact one-port spintronic devices.

{\it Discussions and conclusions.}---Although we have focused on a HH DW as an example, the underlying mechanism is not restricted to this specific configuration. Similar phenomena 
can occur for other types of DWs and spin torques, since their collective coordinate dynamics are governed by equivalent equations after appropriate redefinitions and mappings. In terms of experimental realization, materials with reduced crystal symmetry
or anomalous spin Hall effects \cite{wang2021anomalous} provide a promising platform to generate the required out-of-plane spin polarization, as demonstrated in the recent experiments \cite{safeer2019,li2024}.
In addition, recent observations on ferrimagnetic N{\'e}el DW \cite{song2025,ma2025}
offer another platform to implement our proposal, where the spin polarization $\boldsymbol{p}$ 
 should lie in the film plane and be perpendicular to the DW plane.

In order to evaluate the experimental feasibility of the proposed mechanism, we estimate the current density required.
Considering spin current generated by the conventional spin Hall effect \cite{Sinova2015}, and taking representative
parameters  $\vartheta_{\mathrm{SH},1} \simeq 0.1$, $d_1\simeq1$ nm, 
$ M_1 \simeq 1010 \,\mathrm{kA}/\mathrm{m}$, we obtain a critical current density
$j_{\mathrm{c1}}\simeq 5.4\times10^{11} \mathrm{A}/\mathrm{m}^2$, $j_{\mathrm{c2}}\simeq 4.65\times10^{11} \mathrm{A}/\mathrm{m}^2$ which is achievable
in experiments. Furthermore, our results highlight the role of field-like torque in DW motion, which is usually
underestimated in the conventional approaches. Substantial field-like 
torque has been observed experimentally in various materials 
\cite{wang2022,finley2018,kondou2021,wang2024molybdenum}. Moreover, 
a uniform magnetic field along the $y$-direction 
provides an additional tuning knob to effectively control the field-like torque, 
allowing the potential energy landscape to be adjusted to favor reversible 
DW motion.

In summary, we have demonstrated that inertia can qualitatively alter the
steady DW dynamics in ferrimagnets near the AMCP. Contrary to the conventional expectation that inertia influences only the transient responses, we show that a DW driven by a direct current can
propagate steadily in either direction. The reversible DW motion is well understood from the inertial dynamics
of DW tilting angle, which evolves in a current-induced double-well potential. The DW inertia allows the relaxation of DW toward different stable points with distinguished DW velocities.
Our analytical theory, supported by micromagnetic simulations, identified two critical currents that 
allow the existence of reversible DW motion, which is robust against fine-tuning of magnetic parameters. Beyond its fundamental implications, the bistable DW response enables novel functionalities based on DW motion, including sensitive magnetic field detection and reconfigurable one-port spintronic devices. More broadly, our results establish inertia as a key ingredient in steady magnetization dynamics and open new avenues for exploiting inertial effects in driven-dissipative magnetic systems.

{\it Acknowledgments.}-- This work is supported by the National Key R$\&$D Program of China (2022YFA1402700) and the National Natural Science Foundation of China (NSFC) (Grant No. 12574132). XRW acknowledges the support from the University Development Fund of the 
Chinese University of Hong Kong, Shenzhen, and the Guangdong Provincial 
Quantum Strategy Special Project.


\begin{thebibliography}{99}		

%
%
%
%
%


\bibitem{FIMexper1}
K.-J. Kim, S. K. Kim, Y. Hirata, S.-H. Oh, T. Tono, D.-H. Kim, T. Okuno, W. S. Ham, S. Kim, G. Go, Y. Tserkovnyak, 
A. Tsukamoto, T. Moriyama, K.-J. Lee and T. Ono,
Fast domain wall motion in the vicinity of the angular momentum compensation temperature of ferrimagnets,
\href{https://doi.org/10.1038/nmat4990}{Nat. Mater. \textbf{16}, 1187 (2017)}.



\bibitem{FIMexper6} D.-H. Kim, T. Okuno, S. K. Kim, 
S.-H. Oh, T. Nishimura, Y. Hirata, 
Y. Futakawa, H. Yoshikawa, A. Tsukamoto, Y. Tserkovnyak, 
Y. Shiota, T. Moriyama, 
K.-J. Kim, K.-J. Lee, and T. Ono, Low magnetic 
damping of ferrimagnetic GdFeCo alloys, 
\href{https://journals.aps.org/prl/abstract/10.1103/PhysRevLett.122.127203}
{Phys. Rev. Lett. {\bf 122}, 127203 (2019)}.


\bibitem{ivanov2020} 
B. A. Ivanov, E. G. Galkina, V. E. Kireev, N. E. Kulagin, R. V. Ovcharov, and 
R. S. Khymyn, Nonstationary forced motion of domain walls in 
ferrimagnets near the spin compensation point,
\href{https://doi.org/10.1063/10.0001552}{Low Temp. Phys. {\bf 46}, 841 (2020)}.




	\bibitem{FerriAC}
	M. Jin, I.-S. Hong, D.-H. Kim, K.-J. Lee, and S. K. Kim, 
	Domain-wall motion driven by a rotating field in a ferrimagnet,
	\href{https://journals.aps.org/prb/abstract/10.1103/PhysRevB.104.184431}
	{Phys. Rev. B {\bf 104}, 184431 (2021)}.


	
	\bibitem{FIMexper7} M. V. Logunov, S. S. Safonov, 
A. S. Fedorov, A. A. Danilova, N. V. Moiseev, 
A. R. Safin, S. A. Nikitov, and A. Kirilyuk, Domain 
wall motion across magnetic
and spin compensation points in magnetic garnets, 
\href{https://journals.aps.org/prapplied/abstract/10.1103/PhysRevApplied.15.064024}
{Phys. Rev. Applied {\bf 15}, 064024 (2021)}.	





\bibitem{KYField1}
K. Y. Jing, X. Gong, and X. R. Wang, 
Field-driven domain wall motion in ferrimagnetic nanowires,
\href{https://doi.org/10.1103/PhysRevB.106.174429}
{Phys. Rev. B {\bf 106}, 174429 (2022)}.



\bibitem{yang2015}
S.-H. Yang, K.-S. Ryu, and  S. S. P. Parkin,
Domain-wall velocities of up to 750 m s$^{-1}$ driven by exchange-coupling torque in synthetic antiferromagnets.
\href{https://doi.org/10.1038/nnano.2014.324}{Nat. Nanotechnol., \textbf{10}, 221--226 (2015)}.


\bibitem{FIMexper3} S. A. Siddiqui, J. Han, J. T. Finley, C. A. Ross, 
and L. Liu, Current-induced domain wall motion in a compensated ferrimagnet,
\href{https://journals.aps.org/prl/abstract/10.1103/PhysRevLett.121.057701}
{Phys. Rev. Lett. {\bf 121}, 057701 (2018)}.



\bibitem{FIMexper2} L. Caretta, M. Mann, F. B\"{u}ttner, K. Ueda, B. Pfau, 
C. M. G\"{u}nther, P. Hessing, A. Churikova, C. Klose, M. Schneider, D. Engel, 
C. Marcus, D. Bono, K. Bagschik, S. Eisebitt, and G. S. D. Beach, Fast 
current-driven domain walls and small skyrmions in a compensated ferrimagnet,
\href{https://www.nature.com/articles/s41565-018-0255-3}{Nat. Nanotechnol. {\bf 13}, 1154 (2018)}.

%
%
%


	



\bibitem{FIMexper4}	T. Okuno, D.-H. Kim, S.-H. Oh, S. K. Kim, Y. Hirata, T. Nishimura, 
W. S. Ham, Y. Futakawa, H. Yoshikawa, A. Tsukamoto, 
Y. Tserkovnyak, Y. Shiota, 
T. Moriyama, K.-J. Kim, K.-J. Lee, and T. Ono, 
Spin-transfer torques for domain wall 
motion in antiferromagnetically coupled ferrimagnets,
\href{https://doi.org/10.1038/s41928-019-0303-5}{Nat. Electron. {\bf 2}, 389 (2019)}.  



\bibitem{FIMexper5} E. Haltz, J. Sampaio, 
S. Krishnia, L. Berges, R. Weil, and A. Mougin,
	Measurement of the tilt of a moving domain wall 
	shows precession-free dynamics in 
	compensated ferrimagnets, \href{https://www.nature.com/articles/s41598-020-73049-5}
	{Sci. Rep. {\bf 10}, 16292 (2020)}.
	


\bibitem{FIMexper8}
K. Cai, Z. Zhu, J. M. Lee, R. Mishra, L. Ren, 
S. D. Pollard, P. He, G. Liang, K. L. Teo, H. Yang,
Ultrafast and energy-efficient spin–orbit torque
switching in compensated ferrimagnets, 
\href{https://doi.org/10.1038/s41928-019-0345-8}{Nat. Electron. {\bf 3}, 37 (2020)}.	
	
	
	
		 	\bibitem{Eduardo1}
 	E. Mart\'{i}nez, V. Raposo and \'{O}. Alejos,
 	Novel interpretation of recent experiments on the dynamics of domain 
 	walls along ferrimagnetic strips,
	\href{https://iopscience.iop.org/article/10.1088/1361-648X/aba7eb}
	{J. Phys.: Condens. Matter {\bf 32}, 465803 (2020)}.		
	
	
\bibitem{Ferrisc} V. V. Yurlov, K. A. Zvezdin, P. N. Skirdkov, and A. K. Zvezdin,
Domain wall dynamics of ferrimagnets influenced by spin current near
the angular momentum compensation temperature, 
\href{https://doi.org/10.1103/PhysRevB.103.134442}{Phys. Rev. B {\bf 103}, 134442 (2021)}.


\bibitem{FIMexper9}
S. Ghosh, T. Komori, A. Hallal, J. Pe\~{n}a Garcia, T. Gushi, T. Hirose, H. Mitarai, 
H. Okuno, J. Vogel, M. Chshiev, J.-P. Attan\'{e}, L. Vila, T. Suemasu, and S. Pizzini,
Current-driven domain wall dynamics in ferrimagnetic nickel-doped Mn$_4$N films: 
very large domain wall velocities and reversal of motion direction across the magnetic compensation point,
\href{https://doi.org/10.1021/acs.nanolett.1c00125}{Nano Lett. {\bf 21}, 6, 2580 (2021)}.

	

		\bibitem{FerriCD1}
	J. L. Liu, P. B. He, and M. Q. Cai,
	Current-driven spiral domain wall in a ferrimagnet near the magnetization compensation point,
\href{https://doi.org/10.1103/PhysRevResearch.4.023253}
{Phys. Rev. Research {\bf 4}, 023253 (2022)}.






	

	\bibitem{FerriSOT}
	G. Sala and P. Gambardella,
	Ferrimagnetic dynamics induced by spin-orbit torques,
	\href{https://doi.org/10.1002/admi.202201622}
	{Adv. Mater. Interfaces, {\bf 9}, 2201622 (2022)}.




\bibitem{KYcurrent1}
K. Y. Jing, Z. Z. Sun and X. R. Wang,
Current-driven domain wall motion in ferrimagnetic nanowires,
\href{https://doi.org/10.1103/PhysRevB.110.054414}{Phys. Rev. B \textbf{110}, 054414 (2024)}.


	


\bibitem{yan2011}
P. Yan, X. S. Wang, and X. R. Wang,
All-magnonic spin-transfer torque and domain wall propagation.
\href{https://doi.org/10.1103/PhysRevLett.107.177207}{Phys. Rev. Lett., \textbf{107}, 177207 (2011)}.



	\bibitem{lan2017}
J. Lan, W. Yu, and J. Xiao,
Antiferromagnetic domain wall as spin wave polarizer and retarder.
\href{https://doi.org/10.1038/s41467-017-00265-5}{Nat. Commun., \textbf{8}, 178 (2017)}.

	\bibitem{yu2018} W. Yu, J. Lan, and J. Xiao, Polarization-selective spin wave driven domain-wall motion in antiferromagnets. \href{https://doi.org/10.1103/PhysRevB.98.144422}{Phys. Rev. B, \textbf{98}, 144422 (2018)}.


	\bibitem{FerriSW}
	S.-H. Oh, S. K. Kim, J. Xiao, and K.-J. Lee,
	Bidirectional spin-wave-driven domain wall motion in ferrimagnets,
	\href{https://journals.aps.org/prb/abstract/10.1103/PhysRevB.100.174403}
	{Phys. Rev. B {\bf 100}, 174403 (2019)}.	
	
	
\bibitem{liang2022}
X. Liang, Z. Wang, P. Yan, and Y. Zhou,
Nonreciprocal spin waves in ferrimagnetic domain-wall channels.
\href{https://doi.org/10.1103/PhysRevB.106.224413}{Phys. Rev. B, \textbf{106}, 224413 (2022)}.
	
	

	
	
\bibitem{xiansi2} 
X. S. Wang and X. R. Wang, 
Thermodynamic theory for thermal-gradient-driven domain-wall motion,
\href{https://doi.org/10.1103/PhysRevB.90.014414}{Phys. Rev. B \textbf{90}, 014414 (2014)}.


	
	
		\bibitem{Ferrithermal}
A. Donges, N. Grimm, F. Jakobs, S. Selzer, 
U. Ritzmann, U. Atxitia, and U. Nowak,
Unveiling domain wall dynamics of ferrimagnets in thermal magnon 
currents: Competition of angular momentum transfer and entropic torque
\href{https://doi.org/10.1103/PhysRevResearch.2.013293}
{Phys. Rev. Research {\bf 2}, 013293 (2020)}.



	
		\bibitem{Ferrianisotropy}
	W. H. Li, Z. Jin, D. L. Wen, X. M. Zhang, M. H. Qin, and J. M. Liu, 
	Ultrafast domain wall motion in ferrimagnets induced by magnetic anisotropy gradient, 
	\href{https://journals.aps.org/prb/abstract/10.1103/PhysRevB.101.024414}
	{Phys. Rev. B {\bf 101}, 024414 (2020)}.
	


\bibitem{Parkin1} S. S. P. Parkin, M. Hayashi, L. Thomas, 
Magnetic domain-wall racetrack memory, 
\href{https://www.science.org/doi/10.1126/science.1145799}
{Science, {\bf320}, 190 (2008)}.









%





	
	
%
%

%
%
%






\bibitem{Review1} S. K. Kim, G. S. D. Beach, K.-J. Lee, T. Ono, T. Rasing, 
and H. Yang, Ferrimagnetic spintronics,
\href{https://doi.org/10.1038/s41563-021-01139-4}{Nat. Mater. {\bf 21}, 24 (2022)}.





\bibitem{zhang2023}
Y. Zhang, X. Feng, Z. Zheng, Z. Zhang, K. Lin, X. Sun, G. Wang,
J. Wang, J. Wei, P. Vallobra, Y. He, Z. Wang, L. Chen, K. Zhang,
Y. Xu, and W. Zhao, Ferrimagnets for spintronic devices:
From materials to applications.
\href{https://doi.org/10.1063/5.0104618}{Appl. Phys. Rev., \textbf{10}, 011301 (2023)}.









\bibitem{landau1935}
L. D. Landau and E. M. Lifshitz. On the theory of the dispersion of magnetic permeability in ferromagnetic bodies.
\href{https://doi.org/10.1016/B978-0-08-036364-6.50008-9}{Phys. Z. Sowjetunion, \textbf{8}, 153--169 (1935)}.

 \bibitem{gilbert}
T. L. Gilbert, A phenomenological theory of damping in ferromagnetic materials,
\href{https://ieeexplore.ieee.org/document/1353448}
{IEEE Trans. Magn. {\bf 40}, 3443 (2004)}.

\bibitem{xrw1} 
	X. R. Wang, P. Yan, J. Lu, and C. He,
	Magnetic field driven domain-wall propagation in magnetic nanowires,
	\href{https://doi.org/10.1016/j.aop.2009.05.004}
	{Ann. Phys. (NY), {\bf 324}, 1815 (2009)}.
	
\bibitem{xrw2} 
	X. R. Wang, P. Yan, and J. Lu,
	High-field domain wall propagation velocity in magnetic 
	nanowires,
	\href{https://iopscience.iop.org/article/10.1209/0295-5075/86/67001}
	{{Euro. Phys. Lett.} {\bf 86}, 67001 (2009)}.




\bibitem{neeraj2021}
K. Neeraj, N. Awari, S. Kovalev, D. Polley, N. Z. Hagstr{\"o}m, S. S. P. K. Arekapudi, A. Semisalova, K. Lenz, B. Green, J.-C. Deinert, I. Ilyakov, M. Chen, M. Bawatna, V. Scalera, M. d’Aquino, C. Serpico, O. Hellwig, J.-E. Wegrowe, M. Gensch, and S. Bonetti,
Inertial spin dynamics in ferromagnets.
\href{https://doi.org/10.1038/s41567-020-01040-y}{Nat. Phys., \textbf{17}, 245--250 (2021)}.

\bibitem{quarenta2024}
M. G. Quarenta, M. Tharmalingam, T. Ludwig, H. Y. Yuan, L. Karwacki,
R. C. Verstraten, and R. A. Duine. Bath-induced spin inertia.
\href{https://doi.org/10.1103/PhysRevLett.133.136701}{Phys. Rev. Lett., \textbf{133}, 136701 (2024)}.





\bibitem{yuan2025}
H. Y. Yuan, Using surface plasmons to detect spin inertia.
\href{https://doi.org/10.1103/5dgr-4tkt}{Phys. Rev. B, \textbf{112}, 054438 (2025)}.



\bibitem{kim2017vortex}
S. K. Kim and Y. Tserkovnyak,
Fast vortex oscillations in a ferrimagnetic disk near the angular momentum compensation point.
\href{https://doi.org/10.1063/1.4985577}{Appl. Phys. Lett., \textbf{111}, 032401 (2017)}.


\bibitem{kim2017skyrmion}
S. K. Kim, K.-J. Lee, and Y. Tserkovnyak,
Self-focusing skyrmion racetracks in ferrimagnets.
\href{https://doi.org/10.1103/PhysRevB.95.140404}{Phys. Rev. B, \textbf{95}, 140404(R) (2017)}.

\bibitem{liu2022}
Y. Liu, T. T. Liu, Z. Jin, Z. P. Hou, D. Y. Chen, Z. Fan, M. Zeng, X. B. Lu, X. S. Gao, M. H. Qin, and J.-M. Liu,
Spin-wave-driven skyrmion dynamics in ferrimagnets: Effect of net angular momentum.
\href{https://doi.org/10.1103/PhysRevB.106.064424}{Phys. Rev. B, \textbf{106}, 064424 (2022)}.



\bibitem{tveten2013}
E. G. Tveten, A. Qaiumzadeh, O. A. Tretiakov, and A. Brataas,
Staggered dynamics in antiferromagnets by collective coordinates.
\href{https://doi.org/10.1103/PhysRevLett.110.127208}{Phys. Rev. Lett., \textbf{110}, 127208 (2013)}.



\bibitem{shiino2016}
T. Shiino, S.-H. Oh, P. M. Haney, S.-W. Lee, G. Go, B.-G. Park,
and K.-J. Lee. Antiferromagnetic domain wall motion driven by spin-orbit torques.
\href{https://doi.org/10.1103/PhysRevLett.117.087203}{Phys. Rev. Lett., \textbf{117}, 087203 (2016)}.


	
	
	
	
	 \bibitem{effLLG4}
E. Haltz, S. Krishnia, L. Berges, A. Mougin, and J. Sampaio, Domain wall
 dynamics in antiferromagnetically coupled double-lattice systems, 
 \href{https://journals.aps.org/prb/abstract/10.1103/PhysRevB.103.014444}{Phys. Rev. B {\bf 103}, 014444 (2021)}.
 
 




\bibitem{Song2021}
C. Song, R. Zhang, L. Liao, Y. Zhou, X. Zhou, R. Chen, Y. You, X. Chen and F. Pan,
Spin-orbit torques: Materials, mechanisms, performances, and potential applications,
\href{https://doi.org/10.1016/j.pmatsci.2020.100761}{Prog. Mater. Sci. \textbf{118}, 100761 (2021)}.



\bibitem{Slonczewski} 
J. Slonczewski, Current-driven excitation of magnetic multilayers,
\href{https://doi.org/10.1016/0304-8853(96)00062-5}
{J. Magn. Magn. Mater. \textbf{159}, L1 (1996)}.

\bibitem{berger1996}
L. Berger. Emission of spin waves by a magnetic multilayer traversed by a current.
\href{https://doi.org/10.1103/PhysRevB.54.9353}{Phys. Rev. B, \textbf{54}, 9353 (1996)}.


\bibitem{saidaoui2014}
H. B. M. Saidaoui, A. Manchon, and X. Waintal,
Spin transfer torque in antiferromagnetic spin valves: From clean to disordered regimes.
\href{https://doi.org/10.1103/PhysRevB.89.174430}{Phys. Rev. B, \textbf{89}, 174430 (2014)}.




\bibitem{abert2017}
C. Abert, H. Sepehri-Amin, F. Bruckner, C. Vogler, M. Hayashi, and D. Suess,
Fieldlike and dampinglike spin-transfer torque in magnetic multilayers.
\href{https://doi.org/10.1103/PhysRevApplied.7.054007}{Phys. Rev. Appl., \textbf{7}, 054007 (2017)}.




\bibitem{wang2024}
B. Wang, Y. Guo, X. Qi, B. Zhang, Z. Li, Z. Hu, Q. Wang, and J. Cao.
Large dampinglike torque contribution originating from the orbital Rashba-Edelstein effect at a Pt/CoO interface.
\href{https://doi.org/10.1103/PhysRevB.110.104404}{Phys. Rev. B, \textbf{110}, 104404 (2024)}.

\bibitem{yang2024}
Y. Yang, P. Wang, J. Chen, D. Zhang, C. Pan, S. Hu, T. Wang, W. Yue, C. Chen, W. Jiang, L. Zhu, X. Qiu, Y. Yao, Y. Li, W. Wang, and Y. Jiang.
Orbital torque switching in perpendicularly magnetized materials.
\href{https://doi.org/10.1038/s41467-024-52988-x}{Nat. Commun., \textbf{15}, 8645 (2024)}.


\bibitem{fukami2025}
S. Fukami, K.-J. Lee, and M. Kl{"a}ui.
Challenges and opportunities in orbitronics.
\href{https://doi.org/10.1038/s41567-025-03143-w}{Nat. Phys (2025).} 

	\bibitem{AFM1}
	A. Kamra, R. E. Troncoso, W. Belzig, and A. Brataas, 
	Gilbert damping phenomenology for two-sublattice magnets, 
	\href{https://journals.aps.org/prb/abstract/10.1103/PhysRevB.98.184402}
{Phys. Rev. B {\bf 98}, 184402 (2018)}.

		\bibitem{Yuan1}
	H. Y. Yuan, Q. Liu, K. Xia, Z. Yuan, and X. R. Wang, 
	Proper dissipative torques in antiferromagnetic dynamics, 
	\href{https://iopscience.iop.org/article/10.1209/0295-5075/126/67006}
	{Euro. Phys. Lett. {\bf 126}, 67006 (2019)}.
	








\bibitem{rodriguez2024}
R. Rodriguez, M. Cherkasskii, R. Jiang, R. Mondal, A. Etesamirad, A. Tossounian, B. A. Ivanov, and I. Barsukov,
Spin inertia and auto-oscillations in ferromagnets.
\href{https://doi.org/10.1103/PhysRevLett.132.246701}{Phys. Rev. Lett., \textbf{132}, 246701 (2024)}.



\bibitem{Walker} N. L. Schryer and L. R. Walker, 
The motion of $180^{\circ}$ domain walls in uniform dc magnetic fields,
\href{https://aip.scitation.org/doi/10.1063/1.1663252}
{J. Appl. Phys. {\bf 45}, 5406 (1974)}.


%
%
%
%
%
%
%
%
%
%
%
%


\bibitem{mumax3}
A. Vansteenkiste, J. Leliaert, M. Dvornik, M. Helsen, F. G. Sanchez, 
and B. V. Waeyenberge, The design and verification of MuMax3,  
\href{https://aip.scitation.org/doi/10.1063/1.4899186}
{AIP Advances {\bf 4}, 107133 (2014)}.



\bibitem{parameter1}
I. Radu, K. Vahaplar, C. Stamm, T. Kachel, N. Pontius, H. A. Dürr, 
T. A. Ostler, J. Barker, R. F. L. Evans, R. W. Chantrell, A. Tsukamoto, 
A. Itoh, A. Kirilyuk, Th. Rasing, and A. V. Kimel, 
Transient ferromagnetic-like state mediating ultrafast reversal of antiferromagnetically coupled spins,
\href{https://www.nature.com/articles/nature09901}{Nature {\bf 472}, 205 (2011)}.

\bibitem{parameter3}
M. Ding, S. J. Poon,
Tunable perpendicular magnetic anisotropy in GdFeCo amorphous films,
\href{https://www.sciencedirect.com/science/article/pii/S0304885313001522}
{J. Magn. Magn. Mater. {\bf 339}, 51 (2013)}.



\bibitem{parameter2}
C. E. Patrick, S. Kumar, G. Balakrishnan, R. S. Edwards, M. R. Lees, L. Petit, 
and J. B. Staunton, Calculating the magnetic anisotropy of rare-earth–transition-metal ferrimagnets,
\href{https://doi.org/10.1103/PhysRevLett.120.097202}{Phys. Rev. Lett. {\bf 120}, 097202 (2018)}.


\bibitem{parameter4}
H. Damas, A. Anadon, D. C. Berrocal, J. A. Saenz, J. Bello, A. A. C\'{o}rdova, S. Migot, J. Ghanbaja, O. Copie, M. Hehn, V. Cros, S. P. Watelot, J. C. R. S\'{a}nchez,
Ferrimagnet gdfeco characterization for spin-orbitronics: 
large field-like and damping-like torques,
\href{https://doi.org/10.1002/pssr.202200035}
{Phys. Status Solidi RRL {\bf 16}, 2200035 (2022)}.



\bibitem{Shen2024}
L. Shen and K. Shen,
Skyrmion-based chaotic oscillator driven by a constant current,
\href{https://doi.org/10.1103/PhysRevB.109.014422}
{Phys. Rev. B {\bf 109}, 014422 (2024)}.










\bibitem{wang2021anomalous}
X. R. Wang. Anomalous spin Hall and inverse spin Hall effects in magnetic systems.
\href{https://doi.org/10.1038/s42005-021-00557-9}{Commun. Phys., \textbf{4}, 55 (2021)}.



\bibitem{safeer2019}
C. K. Safeer, N. Ontoso, J. Ingla-Aynés, F. Herling, V. T. Pham, A. Kurzmann, K. Ensslin, A. Chuvilin, I. Robredo, M. G. Vergniory, F. de Juan, L. E. Hueso, M. R. Calvo, and F. Casanova,
Large multidirectional spin-to-charge conversion in low-symmetry semimetal MoTe2 at room temperature.
\href{https://pubs.acs.org/doi/10.1021/acs.nanolett.9b03485}{Nano Lett., \textbf{19}, 8758--8766 (2019)}.





\bibitem{li2024}
D. Li, X.-Y. Liu, X.-G. Ye, Z.-C. Pan, W.-Z. Xu, P.-F. Zhu, A.-Q. Wang, K. Watanabe, T. Taniguchi, and Z.-M. Liao,
Facilitating field-free perpendicular magnetization switching with a Berry curvature dipole in a Weyl semimetal.
\href{https://doi.org/10.1103/PhysRevB.110.L100409}{Phys. Rev. B, \textbf{110}, L100409 (2024)}.


\bibitem{song2025}
Y. Song, S. Huang, D. Bono, J. T. Sadowski, C. A. Ross, and G. S. D. Beach.
N{\'e}el domain walls with bistable chirality in a perpendicularly magnetized ferrimagnetic insulator.
\href{https://www.nature.com/articles/s41467-025-60412-1}{Nat. Commun., \textbf{16}, 5201 (2025)}.


\bibitem{ma2025}
Y. Ma, X. Fang, F. Yan, L. Wang, R. Yao, M. Meng, P. Qin, J. Yang, Z. Liu, Z. Luo, S. Ning, and F. Luo.
Magnetic domain wall energy landscape engineering in a ferrimagnet.
\href{https://doi.org/10.1021/acs.nanolett.4c04916}{Nano Lett., \textbf{25}, 261--267 (2025)}.





\bibitem{Sinova2015}
J. Sinova, S. O. Valenzuela, J. Wunderlich, C. H. Back and T. Jungwirth,
Spin Hall effects,
\href{https://doi.org/10.1103/RevModPhys.87.1213}{Rev. Mod. Phys. \textbf{87}, 1213 (2015)}.




\bibitem{finley2018}
J. Finley, C.-H. Lee, P. Y. Huang, and L. Liu.
Spin-orbit torque switching in a nearly compensated Heusler ferrimagnet.
\href{https://doi.org/10.1002/adma.201805361}{Adv. Mater., \textbf{31}, 1805361 (2018)}.



\bibitem{kondou2021}
K. Kondou, H. Chen, T. Tomita, M. Ikhlas, T. Higo, A. H. MacDonald, S. Nakatsuji, and Y. Otani.
Giant field-like torque by the out-of-plane magnetic spin Hall effect in a topological antiferromagnet.
\href{https://doi.org/10.1038/s41467-021-26453-y}{Nat. Commun., \textbf{12}, 6491 (2021)}.

\bibitem{wang2022}
J. Wang, C. Li, R. Tang, G. Chai, J. Yao, and C. Jiang.
Spin-orbit torque in a single ferrimagnetic GdFeCo layer near the compensation temperature.
\href{https://doi.org/10.1063/5.0078995}{Appl. Phys. Lett., \textbf{120}, 102402 (2022)}.

\bibitem{wang2024molybdenum}
X. Wang, A. Meng, Y. Yao, F. Lin, Y. Bai, X. Ning, B. Li, Y. Zhang, T. Nie, S. Shi, and W. Zhao.
Large field-like spin–orbit torque and enhanced magnetization switching efficiency utilizing amorphous Mo.
\href{https://doi.org/10.1021/acs.nanolett.4c01100}{Nano Lett., \textbf{24}, 6931--6938 (2024)}.







%






%
%
%
%
%


%

%
%
%
%
%
%
%
%
%
%
%
%
%
%
%
%
%
%
%
%
%
%
%
%
%
%
%











%
%
%
%
%
%
%
%
%
%
%

%
%
%
%
%
%

%
%

%
%
%
%
%
%
%
%
%
%
%

%
%
%
%
%
%
%
%
%
%
%
%
%
%

%
%
%
%
%

%
%
%
%
%
%
%
%
%
%
%
%
%


%
%

%
%
%

%
%
%

%
%
%
%
%

%
%
%
%
%

%
%
%
	








%


%
%


	
	
%
%
%

%


\end{thebibliography}
\end{document}